# The Tully–Fisher relation for low surface brightness galaxies – implications for galaxy evolution

M.A. Zwaan,[1] J.M. van der Hulst,[1] W.J.G. de Blok[1] and S.S. McGaugh[2]

[1] *Kapteyn Astronomical Institute, P.O. Box 800, 9700 AV Groningen, The Netherlands*
[2] *Institute of Astronomy, Madingley Road, Cambridge CB3 0HA, UK*

**ABSTRACT**

We present the $B$ band Tully–Fisher relation for Low Surface Brightness (LSB) galaxies. These LSB galaxies follow the same Tully–Fisher relation as normal spiral galaxies. This implies that the mass-to-light ratio ($\mathcal{M}/L$) of LSB galaxies is typically a factor of 2 larger than that of normal galaxies of the same total luminosity and morphological type. Since the dynamical mass of a galaxy is related to the rotational velocity and scale length via $\mathcal{M} \propto V^2 h$, at fixed linewidth LSB galaxies must be twice as large as normal galaxies. This is confirmed by examining the relation between scale length and line width for LSB and normal galaxies. The universal nature of the Tully–Fisher relation can be understood if LSB galaxies are galaxies with low mass surface density, $\bar{\sigma}$. The mass surface density apparently controls the luminosity evolution of a galaxy such as to keep the product $\bar{\sigma} \mathcal{M}/L$ constant.

**Key words:** galaxies: fundamental parameters – galaxies: evolution – galaxies: spiral – galaxies: distances and redshifts – cosmology: distance scale.

## 1 INTRODUCTION

It has become clear in recent years that disk galaxies exhibit a large range in central surface brightness and scale length: the two parameters that describe an exponential disk. Freeman's (1970) result that the central disk surface brightness of spiral galaxies falls within a narrow range of $\mu_B(0) = 21.65 \pm 0.30$ mag/$\square''$, has now been superseded (e.g., de Jong and van der Kruit 1994; Davies 1990). At the faint end of the surface brightness range we find the so-called Low Surface Brightness (LSB) galaxies; galaxies with $\mu_B(0) \approx 23$ mag/$\square''$ or fainter.

The selection effects imposed by the night sky and their low surface brightnesses make LSB galaxies difficult to detect. Hence, they are underrepresented in catalogues and have until recently been studied in less detail than normal, High Surface Brightness (HSB) galaxies, i.e., those galaxies with surface brightnesses approximately equal to the Freeman value.

The Tully–Fisher relation (Tully and Fisher 1977, hereafter T–F), the relation between absolute magnitude and rotational velocity, is well-established for HSB spiral galaxies. In this paper we examine whether LSB galaxies follow the same relationship, despite their much lower central surface brightness (see also Sprayberry et al. 1995). The position of LSB galaxies in the luminosity-line width plane is particularly interesting because it will give information about the mass-to-light ratio ($\mathcal{M}/L$) for these galaxies. This parameter is still relatively uncertain for LSB galaxies since a detailed study of rotation curves has only been performed for one LSB galaxy (Bosma, Athanassoula and van der Hulst 1988).

## 2 DATA

### 2.1 The samples

We selected the data from several samples from the literature, which we briefly discuss here.

*McGaugh* (McGaugh and Bothun 1994) and *de Blok* (de Blok, van der Hulst and Bothun 1994) studied LSB galaxies with central surface brightnesses $\mu_B(0) \approx 23$ mag/$\square''$. These galaxies are selected from the UGC (Nilson 1973) and the LSB catalogue of Schombert et al. (1992, hereafter SBSM). Bulge and disk components are deconvolved from the surface brightness profiles and exponential functions are fit to the disk components. The disk magnitudes are calculated by integrating the surface brightness profiles to infinity. However, the disk magnitudes ignore any contribution of light from the bulge (which is nevertheless small for most LSB galaxies in this sample), and will introduce extra light at larger radii. Fortunately, the data necessary to calculate sky-limited aperture magnitudes are available. We calculated these and used them in our derivation of absolute magnitudes. The inclinations are derived from the photometric data.

*Knezek* (Knezek 1993) investigated giant LSB galaxies with mean blue surface brightnesses fainter than 25 mag/$\square''$. The consequence of this selection criterion is that galaxies are included with very bright central surface brightnesses. From the total sample we selected only those galaxies with central surface brightnesses fainter than 22 mag/$\square''$. We calculated the disk magnitudes from the published structural parameters; aperture magnitudes are

**Table 1.** Properties of samples

| (1) | (2) | (3) | (4) | (5) |
|---|---|---|---|---|
| Sample | $N$ | $\langle W_{R,50}^i \rangle$ | $\langle M_B^{b,i} \rangle$ | $\langle \mu_B(0) \rangle$ |
| McGaugh | 14 | 182 ± 162 | -18.0 ± 1.8 | 23.7 ± 0.7 |
| De Blok | 9 | 174 ± 73 | -17.5 ± 1.1 | 23.8 ± 0.5 |
| Knezek | 19 | 269 ± 141 | -20.0 ± 1.2 | 22.4 ± 0.3 |
| Total | 42 | 220 ± 142 | -18.8 ± 1.9 | 23.1 ± 0.8 |

Notes:
(1) sample name
(2) number of galaxies in sample
(3) mean corrected linewidth measured at 50% of peak intensity in km s$^{-1}$
(4) mean corrected absolute blue magnitude
(5) mean blue central surface brightness

not available. We determined the inclinations from the major-to-minor axis ratios from the UGC. Large errors in inclination corrected linewidth are hence unavoidable.

From these samples, we selected all galaxies with inclinations larger than 30° and for which a linewidth at 50% of the peak intensity is available. The linewidths were obtained from the catalogue by SBSM, from LEDA[*] or from Schneider et al. (1990, 1992). Table 1 presents the characteristics of each sample, including standard deviations.

## 2.2 Corrections

The two parameters which define the T–F relation, the magnitude and the velocity width, need to be corrected for inclination and absorption effects. In the following we briefly describe the corrections which we have applied.

We corrected the magnitudes for Galactic and internal extinction to face-on orientation, as outlined in Tully and Fouqué (1985, hereafter TFq). The optical depth $\tau$ and the parameter $f$, which describes to what extent the obscuring matter is mixed with the light, are not well determined for LSB galaxies. Hence, we adopt the standard values derived by TFq for normal spiral galaxies: $\tau = 0.55$ and $f = 0.25$.

We corrected the line widths for random motion effects and inclination as outlined in TFq. The parameter which describes the importance of random motion is not well known for LSB galaxies, so once again the value for HSB galaxies is used: $W_t = 14 \,\mathrm{km\,s^{-1}}$.

[*] Lyon-Meudon Extragalactic Database

L4  *M.A. Zwaan, J.M. van der Hulst, W.J.G. de Blok and S.S. McGaugh*

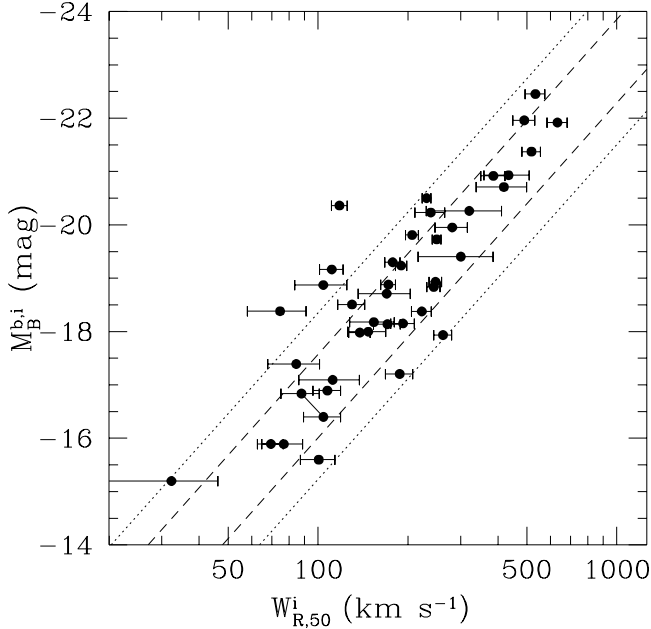

**Figure 1.** Tully–Fisher relation for the sample of LSB galaxies. The long and short dashes represent the $1\sigma$ and $2\sigma$ range of the fit to the Broeils sample. One galaxy that enters the sample twice is connected.

For determining the distances to the galaxies we used a Hubble constant of 75 km s$^{-1}$ Mpc$^{-1}$. We corrected the redshifts for Galactic rotation and a Virgocentric flow of 300 km s$^{-1}$.

## 3 RESULTS

In Fig. 1 we present the $B$ band T–F relation for the sample of LSB galaxies. As a comparison we use a sample of field HSB galaxies compiled by Broeils (1992). We used the same corrections as described in section 2.2. The slope of the T–F relation for these galaxies is $-6.59$, which is in good agreement with slopes found for cluster samples (e.g., Pierce and Tully 1988). The dispersion for the Broeils sample is 0.77 mag. This is considerably larger than what is found for cluster samples, probably due to uncertainties in distances. The $1\sigma$ and $2\sigma$ ranges around the fit to the Broeils sample are represented in Fig. 1 by long and short dashes, respectively. Obviously, the LSB galaxies are indistinguishable from normal galaxies in the luminosity-line width plane, that is, *the T–F relation for LSB galaxies is identical to that for HSB galaxies.*

### 3.1  Morphological type dependence

For normal spiral galaxies, different morphological types have different positions in the luminosity-line width plane (e.g., Kraan-Korteweg, Cameron and Tammann 1988). In order to test whether LSB galaxies follow a similar trend, we made T–F plots binned according to morphological type, both for the LSB and the HSB sample. These are shown in Fig. 2. Only those LSB galaxies which have been unambiguously classified have been plotted. From this figure, it is clear that LSB galaxies follow the same trend with morphological type as the HSB galaxies.

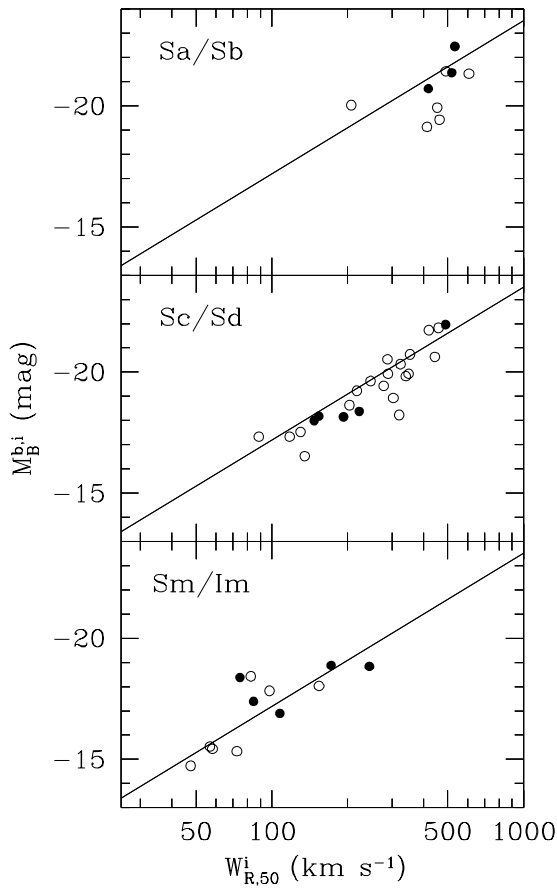

**Figure 2.** Tully–Fisher relations binned according to morphological type. The solid circles are LSB galaxies, the open circles are HSB galaxies from Broeils. The line is a fit to all LSB galaxies.

## 4   DISCUSSION

We have shown that LSB and HSB galaxies obey the same, apparently *fundamental*, T–F relation. Below we discuss the consequences of this result.

### 4.1   Mass-to-light ratios

The fact that LSB galaxies follow a normal T–F relation has important implications for the $\mathcal{M}/L$ of these galaxies. We derive a simple relation between the luminosity, line width, central surface brightness and $\mathcal{M}/L$ to illustrate this. The mass $\mathcal{M}$ is proportional to $V_{\max}^2 h$ and the total luminosity $L$ is proportional to $\Sigma_0 h^2$, where $\Sigma_0$ is the central surface brightness, $h$ the scale length of the disk and $V_{\max}$ the maximum rotational velocity. From these two relations it follows that

$$V_{\max}^4 \propto \frac{\mathcal{M}^2}{h^2} \propto \frac{\mathcal{M}^2 \Sigma_0}{L}, \qquad (1)$$

so that

$$L \propto \frac{V_{\max}^4}{\Sigma_0 (\mathcal{M}/L)^2}, \qquad (2)$$

where one recognises the T–F relation, $L \propto V_{\max}^4$. However, LSB and HSB galaxies are observed to have similar luminosities at a fixed line width, in spite of the difference in $\Sigma_0$. This requires that the difference in surface brightness be compensated by a difference in $\mathcal{M}/L$ so as to keep the product $\Sigma_0 (\mathcal{M}/L)^2$ constant. The mean central surface brightness of our sample is $\langle \mu_B(0) \rangle = 23.1$, i.e., 1.5 mag fainter than typical of the Broeils sample. In order to account for this factor 4 difference in $\Sigma_0$, $\mathcal{M}/L$ must be a factor of 2 *greater* than for normal spiral galaxies of similar total luminosity. This is in good agreement with the typical values of $\mathcal{M}/L$ for LSB galaxies found by van der Hulst et al. (1993).

### 4.2 Scale lengths

The *normal* T–F relation for LSB galaxies and the implication for $\mathcal{M}/L$ should be considered further. Since $\mathcal{M} \propto V_{\max}^2 h$ at any given location in the T–F diagram (i.e., at fixed luminosity and line width), $\mathcal{M}/L$ depends solely on $h$. Therefore, LSB galaxies which have a higher $\mathcal{M}/L$ must also have scale lengths larger by a factor of 2 than those of the HSB counterparts with the same line width. The scale length is the preferred parameter for making a fair comparison between the sizes, as $D_{25}$ severely underestimates the size of a LSB galaxy. The values of the scale lengths of the Broeils sample can be derived from $D_{25}$ which is related to the scale length via $D_{25} = 6.17\,h$ assuming that these galaxies can be represented by exponential disks with central surface brightnesses of 21.65 mag/$\square''$.

In Fig. 3 we show a plot of the scale length of LSB and HSB galaxies versus line width. A strong relationship between scale length and line width exists and there is a significant segregation between the LSB and HSB galaxies. Although a minor difference between the slopes is present, both groups of galaxies follow a similar trend. In order to derive a reliable value for the offset between the two samples, we made double regression fits, requiring the slopes to be equal. The difference between the intercepts of the fits is 0.35 dex, which means that LSB galaxies have scale lengths which are 2.2 times larger than those of normal galaxies with the same line width. This is in good agreement with the factor of 2 found for the difference in $\mathcal{M}/L$ for this sample of galaxies.

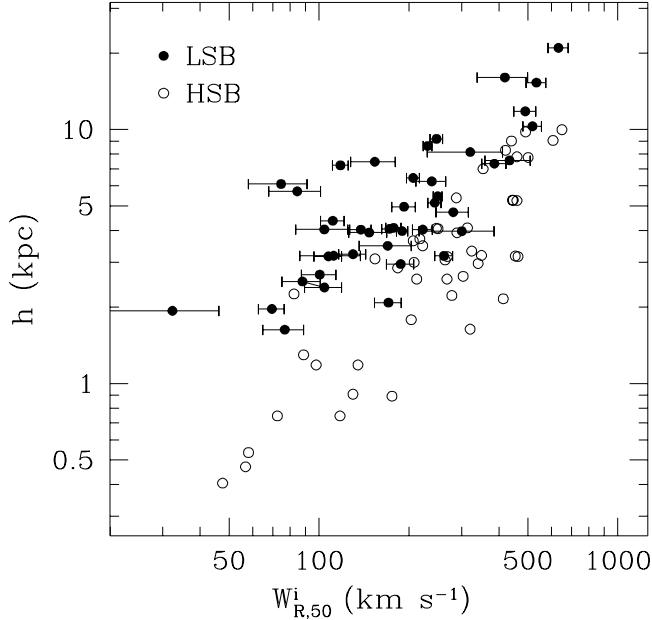

**Figure 3.** Scale length versus velocity width, the solid circles are LSB galaxies, the open circles are HSB galaxies from Broeils.

### 4.3 Mass surface density

In addition to comparing $\mathcal{M}/L$ ratios and scale lengths at fixed line width, one can consider the implications of a universal T–F relation for the mass surface density. From $L \propto V_{\max}^4$ and $V_{\max}^2 \propto \mathcal{M}/h$ it follows that $L \propto \mathcal{M}^2/h^2$, or that

$$\mathcal{M}/L \propto h^2/\mathcal{M} = 1/\bar{\sigma}, \tag{3}$$

where $\bar{\sigma}$ is the average mass surface density. In other words, galaxies with similar mass surface densities have similar $\mathcal{M}/L$ ratios, i.e., the product $\bar{\sigma}\,\mathcal{M}/L$ is constant. The result that LSB galaxies have higher $\mathcal{M}/L$ ratios than HSB galaxies with the same rotational velocity implies that LSB galaxies have lower mass surface densities. Another way of phrasing this result is to combine Eq. (2) and (3) to $\Sigma_0 \propto \bar{\sigma}^2$; the surface brightness of a galaxy apparently is determined by its mass surface density, i.e., LSB galaxies must be less dense than HSB galaxies. The implication is that the mass surface density may be an important parameter controlling the luminosity evolution of a galaxy. Less dense galaxies evolve more slowly, forming relatively fewer stars in a Hubble time, which results in lower surface brightness disks and higher $\mathcal{M}/L$ ratios. This makes sense given that the stability of a disk, and hence the star formation activity, is primarily controlled by the mass surface density (e.g., Toomre 1964; Goldreich and Lynden-Bell 1965; Kennicutt 1989; van der Hulst et al. 1987, 1993). In this light, the T–F relation simply describes the stability of galaxies.

## 5 CONCLUSIONS

We have shown that the $B$ band Tully–Fisher relation for LSB field galaxies is indistinguishable from the relation for their high surface brightness counterparts. This implies that the Tully–Fisher relation is rather fundamental to spiral disks. Basic theoretical considerations require that LSB galaxies have $\mathcal{M}/L$ ratios which are a factor of 2 larger than those of normal spiral galaxies of comparable total luminosity and morphological type. This difference implies that at a fixed rotational velocity, LSB galaxies should have twice as large scale lengths as normal galaxies. This is confirmed by the data. The universal nature of the Tully–Fisher relation requires that galaxies of similar mass surface density have similar $\mathcal{M}/L$ ratios and it is suggestive that the mass surface density of galaxies is a crucial parameter controlling their luminosity evolution.

## ACKNOWLEDGEMENTS

We thank T.S. van Albada for helpful comments on this paper.

This paper has been produced using the Blackwell Scientific Publications LATEX style file.